\documentclass[10pt]{article}
\usepackage{graphicx}
\usepackage[T1]{fontenc}
\usepackage[utf8]{inputenc}
\usepackage{amssymb}
\usepackage{amsfonts}
\usepackage{dsfont}
\usepackage{mathtools}
\usepackage{amsthm}
\usepackage{amsmath}
\usepackage{relsize}
\usepackage{textcomp}
\usepackage{eurosym}
\usepackage{stmaryrd}
\usepackage{xcolor}
\usepackage[multiple]{footmisc}
\usepackage{bigints}
\usepackage{geometry}
\def \N{\mathbb{N}}

\def \Q{\mathbb{Q}}
\def \R{\mathbb{R}}

\def\Fc{{\cal F}}

\def\Nc{{\cal N}}

\def\Tc{{\cal T}}

\def\Vc{{\cal V}}

\def\d{\mathrm{d}}

\begin{document}

\title{Dispensing with optimal control: a new approach for the pricing and management of share buyback contracts}

\author{Bastien \textsc{Baldacci}\footnote{HSBC France, 38 avenue Kléber, 75116 Paris, France.} \and Philippe \textsc{Bergault}\footnote{Université Paris Dauphine-PSL, Ceremade, Place du Maréchal de Lattre de Tassigny, 75116 Paris, France, bergault@ceremade.dauphine.fr.} \and Olivier \textsc{Guéant}\footnote{Université Paris 1 Panthéon-Sorbonne, UFR 27 Mathématiques et Informatique, Centre d'Economie de la Sorbonne, Paris, France, olivier.gueant@univ-paris1.fr.}}
\date{}

\maketitle
\setlength\parindent{0pt}

\begin{abstract}

This paper introduces a novel methodology for the pricing and management of share buyback contracts, overcoming the limitations of traditional optimal control methods, which frequently encounter difficulties with high-dimensional state spaces and the intricacies of selecting appropriate risk penalty or risk aversion parameter. Our methodology applies optimized heuristic strategies to maximize the contract's value. The computation of this value utilizes classical methods typically used for pricing path-dependent options. Additionally, our approach naturally leads to the formulation of a $\Delta$-hedging strategy and disentangles therefore the repurchase strategy from the hedging of the payoff. \\

\vspace{5mm}

\end{abstract}

\section*{Introduction}

The analysis of payout policies, especially the distinction between dividends and share buybacks, is fundamental to corporate finance. Modigliani and Miller's theorem (see \cite{modigliani1958cost}), which asserted the irrelevance of payout policies in a perfect market, profoundly influenced this topic. Serving as a benchmark, this seminal result suggests that payout policy considerations only become relevant in the presence of market imperfections, including taxes, transaction costs, and informational asymmetry. These imperfections have a significant impact on a firm’s payout strategy. Dividends are often viewed as indicators of consistent earnings and a commitment to future stability, attracting investors who seek regular and dependable income.  Beyond their flexibility, share buybacks are strategically favored for several reasons, including signaling stock undervaluation and deterring potential takeovers. They are also appealing to companies aiming to recalibrate their capital structure. Each mechanism meets specific strategic requirements, reflecting a firm's operational situation and market perception (see for instance \cite{allen2003payout}).\\

Over the past decades, share buybacks have globally surged, even outpacing dividend increases, though they faced a downturn in 2023 in the US due to rising interest rates, as US companies frequently finance buybacks with borrowed funds. The shift from fixed-price tender offers and Dutch auctions to predominantly open market repurchases (OMRs) in the 90s, followed by the rise of Accelerated Share Repurchase (ASR) contracts and buyback mandates, historically marks a significant evolution in how buybacks are conducted. ASR setups are more common in the US. This approach involves a company rapidly repurchasing its own shares through a contract with an investment bank. Typically, the bank initially borrows the shares to deliver them to the company and then gradually closes its short position by buying shares in the open market. In contrast, the mandate format, which is more prevalent in Europe, allows the bank to slowly purchase shares on the market and progressively deliver them to the company.\\

In both ASR contracts and share buyback mandates, the bank's compensation is often tied to a price benchmark, typically the Volume Weighted Average Price (VWAP) over a period decided dynamically by the bank itself, which thus faces an optimal stopping time problem. Competitive pricing strategies in these contracts hinge on offering discounts relative to this benchmark. A company may request specific features in these contracts, and banks, particularly those advising the firm, can propose complex terms. Recent years have seen an increase in original features within contracts such as floating notional, lookback features, day exclusion upon insufficient daily volume or high price increase and subsequent maturity extensions or notional reduction. This trend emphasizes the importance of quantitative analysts in accurately pricing and managing these contracts.\\

Often seen as variants of optimal execution problems, the pricing and management of share buyback contracts have mainly been addressed with the tools of stochastic optimal control. Jaimungal and his coauthors in \cite{jaimungal2013optimal}, followed by Guéant and his coauthors in \cite{gueant2015accelerated}, and then Guéant in \cite{gueant2017optimal} have indeed proposed several models relying on optimal control tools for both fixed quantity and fixed notional ASR contracts. These models ultimately involve solving Hamilton-Jacobi-Bellman equations, or simply Bellman equations, using grid or sometimes tree methods. While these methods are efficient in simple cases and have been adopted by several banks, they encounter three significant challenges. First, dealing with complex contracts or advanced price dynamics necessitates a high-dimensional state space, with mandates posing greater difficulties in this regard than ASR contracts. Although neural networks can potentially overcome the curse of dimensionality (see \cite{gueant2020accelerated} and \cite{hamdouche2022policy} for techniques inspired by reinforcement learning ideas), their opaque decision-making process and unpredictable behavior with extreme state values often lead to their rejection by practitioners. The second issue concerns the selection of an appropriate risk penalty and/or risk aversion parameter to mitigate contract execution risks. This choice significantly influences both optimal strategies and pricing. Moreover, the concept of pricing tied to optimal control methods is indifference pricing, a concept seldom used by practitioners. Third, and related to the previous points, optimal control tools often conflict with the pricing and hedging frameworks that are commonly found in the libraries of most investment banks.\\

To address these limitations, Baldacci \textit{et al.} introduced an alternative approach in \cite{baldacci2023new} that simplifies the problem through a heuristic repurchase strategy. This strategy intuitively aligns with the expected directionality of parameter effects, reducing the complexity to the more manageable task of pricing a path-dependent option with American or Bermudean features. Our research builds upon this foundation by refining and optimizing heuristic strategies, thereby providing more detailed insights into the management of share buyback contracts. Specifically, we maximize the expected value under $\mathbb{Q}$ of the payoff associated with heuristic repurchase and stopping strategies using state-of-the-art hyperparameter optimization tools. The resulting payoff can then be hedged dynamically using standard $\Delta$-hedging techniques available in conventional pricing libraries. Notably, our approach disentangles the repurchase strategy from the $\Delta$ of the payoff.

\section*{Stochastic optimal control versus heuristic strategies}

\subsection*{A simple framework}

In practice, the problem faced by a trader entering a buyback contract is a discrete one: each day, the trader chooses the quantity to be bought. In this section, in order to simplify the mathematical presentation, we consider a continuous-time approximation of the problem, as often done in the literature. \\

Assuming zero interest rates for the sake of simplicity, we consider that the price process $(S_t)_t$ of the stock is given by
$$S_t = S_0 + \int_0^t\sigma S_u \d W_u,$$
where $S_0, \sigma>0$ are given and $(W_t)_t$ is a standard Brownian motion under the risk-neutral probability $\Q$. Then the process $(A_t)_t$ corresponding to the time-weighted average price process has the following dynamics:
$$\d A_t = \frac{S_t - A_t}{t} \d t.$$

In our first and simple framework, we consider the case of a theoretical buyback contract with notional $F>0$, maturity $T>0$, and no constraint whatsoever. We assume that there is no friction in the market, i.e. the trader can buy or sell any number of shares at any time without transaction costs or market impact.\footnote{We diverge here from the traditional perspective commonly found in the academic literature, which treats buyback issues primarily as execution challenges. This departure is partly based on the idea that the market has already assimilated the informational content of trades. Moreover, we posit that the option component inherent in buyback contracts holds more significance for most assets than the aspects related to execution costs.}\\

If the contract states that the bank gets paid the time-weighted average price upon delivery of the shares at time $\tau$ for an amount $F$ spent on the market, then the payoff of the bank is $F \frac{A_\tau}{S_\tau} - F$ and the bank simply has to solve the following maximization problem:
\begin{align*}
    \underset{\tau \in \Tc_{0,T}}{\sup} \mathbb E^\mathbb Q \left[ \frac{A_\tau}{S_\tau}\right],
\end{align*}
where for all $t\in [0,T]$, $\Tc_{t,T}$ is the set of stopping times taking values in $[t,T]$.\\

Defining the process $(Y_t)_t$ by $Y_t = \frac{A_t}{S_t}, \forall t \in (0,T]$ and $Y_0 = 1$, we clearly have that, for $t>0$,
\begin{align}\label{dynamX}
    \d Y_t = \left( \frac{1-Y_t}{t} + \sigma^2 Y_t \right) \d t - \sigma Y_t \d W_t.
\end{align}
In particular, $(Y_t)_t$ is Markovian, and our problem boils down to
\begin{align}\label{optstop0}
    \underset{\tau \in \Tc_{0,T}}{\sup} \mathbb E^\mathbb Q \left[Y_\tau \right].
\end{align}

\subsection*{The PDE approach}

Problem \eqref{optstop0} is an optimal stopping problem that can be solved using standard tools from stochastic optimal control theory. Let us introduce the value function
\begin{align}
    u : (t,y) \in [0,T]\times \R_+^*  \mapsto \underset{\tau \in \Tc_{t,T}}{\sup} \mathbb E^\mathbb Q \left[ Y_\tau \bigg| Y_t = y \right].
\end{align}

It is then well known that $u$ satisfies the following quasi-variational inequality in the viscosity sense:
\begin{align}\label{VI0}
\begin{cases}
    \min \left\{-\partial_t u(t,y) - \frac 12 \sigma^2 y^2 \partial^2_{yy} u(t,y) - \left(\sigma^2 y + \frac{1-y}{t} \right) \partial_y u(t,y), u(t,y) - y\right\} = 0\quad \forall (t,y) \in [0,T)\times \R_+^* ,\\
    u(T,y) = y \quad \forall y\in \R_+^*.
\end{cases}    
\end{align}
Of course, this equation cannot be solved in closed-form. In order to obtain the pricing function and optimal strategy associated with Problem \eqref{optstop0}, a classical method consists in using an implicit Euler scheme to solve Equation \eqref{VI0} on a grid. Such schemes are widely used in practice, but they require to introduce some boundary conditions for the numerical computation, such as Neumann conditions, and there is of course a discretization error.

\subsection*{The Longstaff-Schwartz approach}

For $N\in \N^*$, let us introduce a subdivision $\Delta^N = \{0=t_0 <t_1 < \ldots < t_N=T\}$ of the interval $[0,T]$.\\

The dynamic programming principle associated with the discretization of the problem over the above subdivision is written as follows:
\begin{align}\label{DPP}
    u\left(t_n, y \right) = \max \left\{ \mathbb E^\Q \left[u\left(t_{n+1}, Y_{t_n+1} \right) \bigg| Y_{t_n}=y \right], y \right\}, \quad \forall (n,y) \in \{0, \ldots, N-1\}\times \R_+^*.
\end{align}

The Longstaff-Schwartz method is a Monte Carlo method based on the above dynamic programming principle. One first needs to simulate $M \in \N^*$ sample paths of the process $\left(Y^1_{t_n} \right)_n, \ldots,\left(Y^M_{t_n} \right)_n $. For each $m \in \{1, \ldots, M\}$, we have the following terminal condition:
$$u\left(t_N, Y^m_{t_N}\right) = Y^m_{t_N}.$$
In order to proceed by backward induction, one needs to compute at each time $n \in \{0, \ldots, N-1\}$ and for each sample path $m$ the value of
$$\mathbb E^\Q \left[u\left(t_{n+1}, Y_{t_{n+1}}\right) \bigg| Y_{t_n}=Y^m_{t_n} \right].$$
The idea of Longstaff and Schwartz \cite{longstaff2001valuing} consists in approximating the above conditional expectation by regressing at each time step the values $\Big(u \big(t_{n+1},  Y^m_{t_{n+1}}\big)\Big)_m $ on simple functions of $(Y^m_{t_n})_m$. This method can be very effective, but highly depends on the choice of the basis of functions used in the regression.

\subsection*{An optimized heuristic strategy}

Instead of choosing boundary conditions for Equation \eqref{VI0}, or a basis for the regression in order to approximate the conditional expectation in Equation \eqref{DPP}, one can wonder if it is possible to choose directly a parametric form for the execution frontier and then optimize over the parameters.\\ 

More precisely, the idea consists first in choosing a family of functions
$$\Fc := \left\{ f(\cdot,\cdot;\beta) : [0,T]\times \R_+^* \rightarrow \mathbb R \left| \beta \in B \right. \right\},$$
where $B\subset \R^d.$ For each $\beta \in B$, we define the stopping time
$$\tau^\beta = \inf \left\{ t\in [0,T] \left| f(t, Y_t ; \beta) >0 \right. \right\},$$
with the usual convention $\inf \emptyset = T$. We can then approximate the value of $\mathbb E^\Q \left[ Y_{\tau^\beta} \right]$ with a standard Monte Carlo method, and then use state-of-the-art algorithms\footnote{In the numerical examples below we used the Quasi-Monte Carlo (QMC) and the Tree-structured Parzen Estimator (TPE) samplers of the \texttt{optuna} python library.} in order to optimize that value over $\beta \in B$. If the family $\Fc$ is rich enough, the approximate problem
\begin{align}\label{optstop0approx}
    \underset{\beta \in B}{\sup} \mathbb E^\Q \left[ Y_{\tau^\beta} \right]
\end{align}
should be close to Problem \eqref{optstop0}.

\subsection*{Comparison of the different methods}

In this part, we compare numerically the performance of the strategies obtained with the three methods described above. We consider the following parameters:
\begin{itemize}
    \item Initial price $S_0 = 10$ \euro;
    \item Volatility $\sigma = 0.2\ \text{year}^{-1/2}$;
    \item Time horizon $T = \frac 1{12}\ \text{year}$.\\
\end{itemize}

For the first approach, hereafter denoted by PDE, we use an implicit Euler scheme for Equation \eqref{VI0} on $[0,T] \times \left[0.8, 1.2\right]$ with Neumann conditions at the boundaries.\\

For the second approach, hereafter denoted by LS, we approximate the conditional expectations with polynomials of degree $2$ in the $y$ variable.\\

For the third approach, hereafter denoted by OHS, the family $\Fc$ is given by
$$\Fc := \left\{ f(\cdot, \cdot;\beta) : [0,T]\times \R_+^* \rightarrow \mathbb R \left| \beta \in \R^3 \right. \right\},$$
where for all $(t,y)\in [0,T]\times \R_+^* $ and $\beta \in \R^3$, 
$$f\left(t,y;\beta \right) = \beta_1 + \beta_2 \left( 1 - \frac tT \right) + \beta_3 \left( 1 - \frac tT \right)^2 - (y-1). $$

\begin{figure}[h!]
\centering
\includegraphics[width=\textwidth]{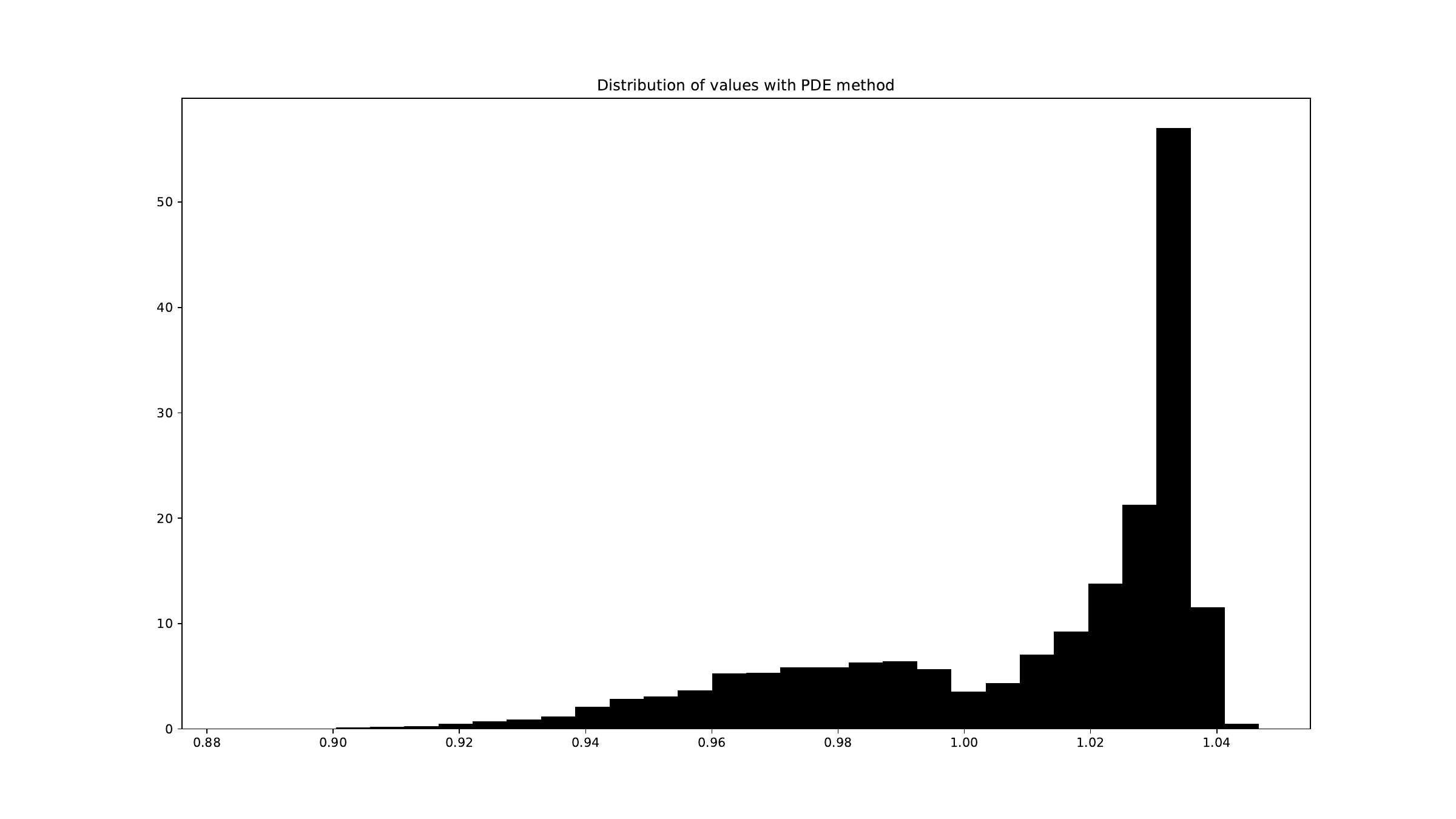}\\
\caption{Distribution of the values obtained with the PDE method.}
\label{distPDE}
\end{figure}

\begin{figure}[h!]
\centering
\includegraphics[width=\textwidth]{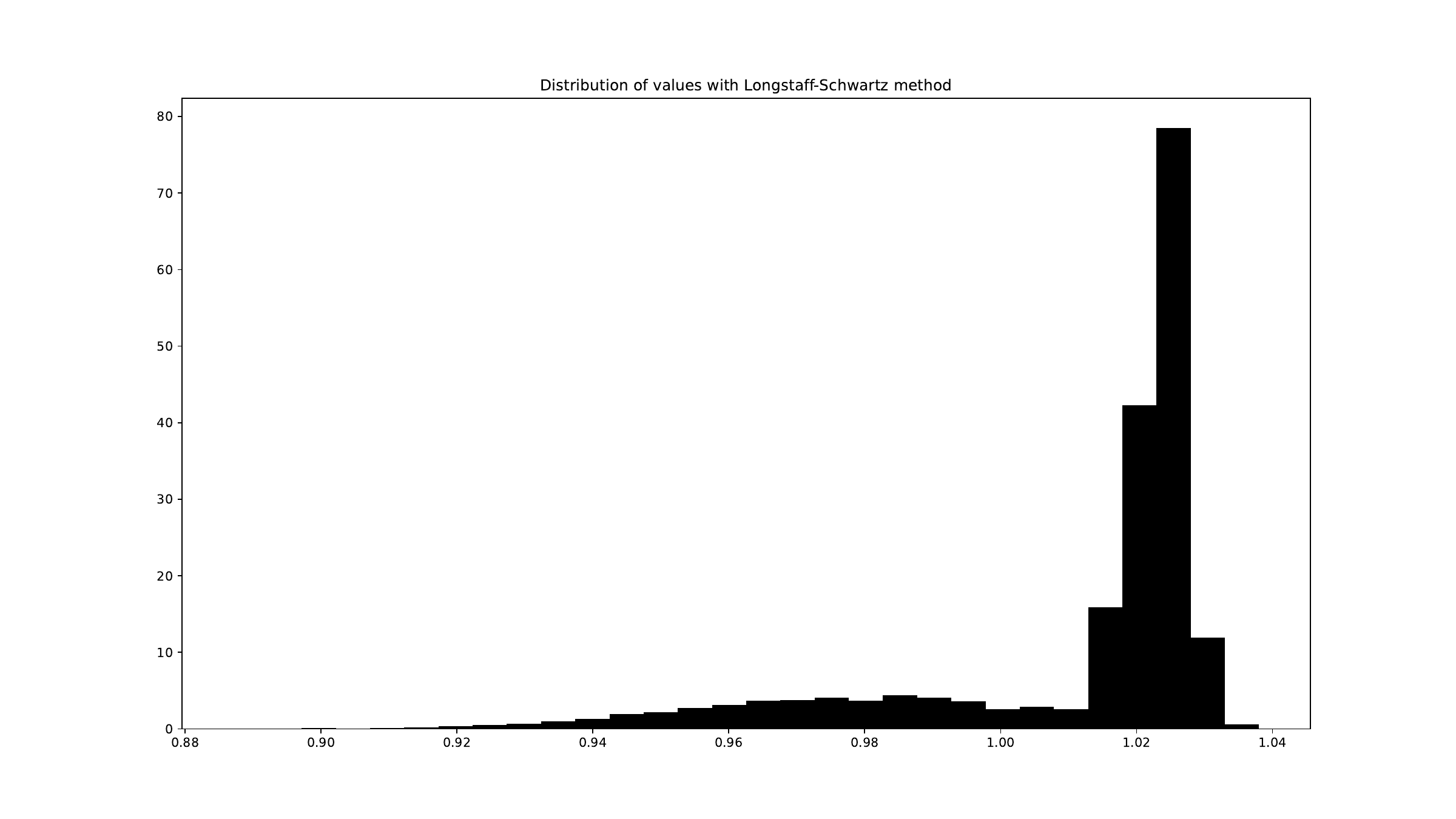}\\
\caption{Distribution of the values obtained with the LS method.}
\label{distLS}
\end{figure}

\begin{figure}[h!]
\centering
\includegraphics[width=\textwidth]{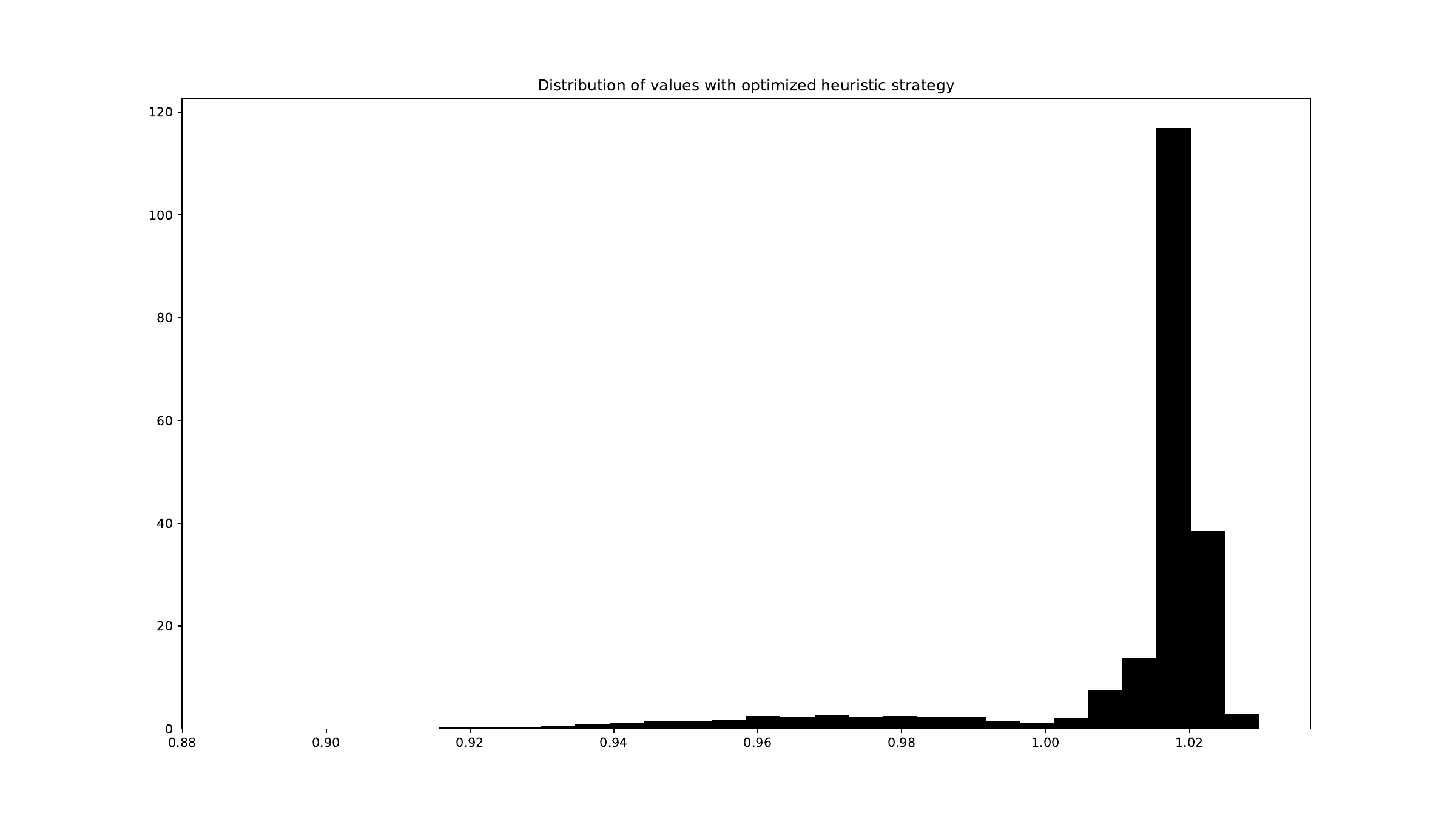}\\
\caption{Distribution of the values obtained with the OHS method.}
\label{distOHS}
\end{figure}

We then test the strategies obtained using the three different methods through a Monte Carlo simulation with 10,000 trajectories. The average payoff and standard deviation under $\mathbb{Q}$ for each method are reported in Table \ref{AStable}. Remarkably, the three strategies yield very similar results in terms of both expected value and standard deviation. The distributions under $\mathbb{Q}$ of the payoffs resulting from the different methods are plotted in Figures \ref{distPDE}, \ref{distLS}, and \ref{distOHS}. The portion of the distributions below 1 is quite spread out, while a spike is observed above 1 in the three cases. Naturally, the risk of the payoff can be hedged away using standard $\Delta$-hedging, and that is why we focused on the maximization of the expected value under $\mathbb{Q}$.

\begin{table}[h!]
\vspace{7mm}
\begin{center}
\begin{tabular}{c | c c c } 
 \hline
Method & PDE & LS & OHS  \\ [1ex] 
\hline
Expected value & 1.0099 & 1.0112 & 1.0111\\ [0.6ex]  
 \hline
Standard deviation & 0.0298 & 0.0241 & 0.0194 \\ [0.6ex] 
\hline
\end{tabular}
\end{center}
\caption {Results for the different strategies.}
\label{AStable}
\end{table}

\newpage

\newpage

\section*{Back to buybacks}

\subsection*{A high-dimensional problem}

The three methods -- PDE, Longstaff-Schwartz, and Optimized Heuristic Strategy -- yield very similar results in terms of pricing for the simple contract discussed in the previous section. However, although the PDE and Longstaff-Schwartz methods can be easily implemented in the simple setting we considered earlier, they become impractical for more complex buyback contracts with numerous features. This impracticality arises because both methods suffer from the curse of dimensionality.\\

Many buyback contracts incorporate indeed constraints related to the daily volume that can be executed, which may vary based on factors such as the current asset price or the total market traded volume. Additionally, the computation of the average price may exclude certain days, for instance, if the trading volume is exceptionally low or the asset price is particularly high.\\

In what follows we consider a buyback contract with a floating maturity comprised between $N_{min}>0$ and $N_{max}\ge N_{min}$, and a floating notional comprised between $F_{min}>0$ and $F_{max}\ge F_{min}$. The trader can choose to exercise early after a given date $N_{ex}$, with $0\le N_{ex}\le N_{min}$.\\

We consider here a discrete-time model with daily decision making: on day~$n$, the trader chooses the quantity $v_n \in [v_{min}, v_{max}]$ to be bought, with $0\le v_{min}< v_{max}$ specified in the contract.\\

There may be a cap $S_{max}$ on the price such that, if $S_n>S_{max}$, the day is suspended, i.e. excluded from the computation of the average price.\footnote{A barrier $V_{min}$ might also be enforced: if the total market volume $V_n$ on day $n$ is such that $V_n< V_{min}$, the day $n$ is suspended. However, this case is not treated in this paper, because the variability of $V_n$ is not modeled and can hardly be hedged.} For all $n$, let us denote by $C_n$ the number of suspended days up to time $n$. Initially, the maturity of the contract is given by $N_{min}$. When a trading day is suspended, the maturity of the program is extended by one day up to the maximal maturity $N_{max}$. If the number of suspended days exceeds $N_{max} - N_{min}$, $F_{min}$ is reduced pro-rata by the excess number of suspended days, i.e.
$$F_{min,n} := F_{min}\left( 1- \frac{\max \left(C_n - (N_{max}-N_{min}), 0 \right)}{N_{max}} \right).$$

\subsection*{Buybacks as payoffs}

Let us denote by $q_n$ the number of shares bought by the trader and delivered to the company until day $n$, and $X_n$ the associated cash spent by the trader up to day $n$. We have $q_0 = X_0 = 0$, and
\begin{align}
    \begin{cases}
        q_{n+1} = q_n + v_n, \\
        X_{n+1} = X_n + v_n S_n,
    \end{cases}
\end{align}
for each $n \ge 0$.\footnote{We impose that the cash spent remains below the maximum notional $F_{max}$ by forbidding trades once it is reached.} Moreover, $S_0$ is known, and we assume that
\begin{align*}
    S_{n+1} = S_n e^{\bar \sigma  \xi_{n+1}},
\end{align*}
where $\bar \sigma>0$ is the daily volatility of the price and $(\xi_n)_n$ are i.i.d. $\Nc (0,1)$ random variables.\\

For each $n\ge 0$, let us denote by $P_n$ the set of days that were not suspended up to day $n$, i.e. 
$$P_n = \left\{ k \in \{0, \ldots, n\} | \text{ day $k$ was not suspended}\right\}.$$

We also denote by $\#P_n$ the cardinal of $P_n$. We then introduce the process $(A_n)_n$ for the cumulative average of the price process $(S_n)_n$ excluding the suspended days, i.e.
$$A_n = \frac 1{\#P_n} \underset{k \in P_n}{\sum} S_k.$$

In contrast to the simpler case discussed previously, it is not immediately evident that the buyback contract incorporating additional features can still be formulated as a payoff structure. Nevertheless, for a predetermined strategy $(v_n)_n$, it becomes apparent that if the trader decides to stop at time $\tilde{N} \ge N_{ex}$ (after $N_{ex}$, the trader has the right to stop whenever $F_{min,\tilde N} \le X_{\tilde N} \le F_{max}$), the resulting payoff can be precisely expressed as
\begin{align*}
    q_{\tilde N} A_{\tilde N} - X_{\tilde N}.
\end{align*}

To incentivize the trader to spend at least the minimum required notional,\footnote{At the stopping time, we buy the required number of shares to reach the minimum notional if it has not been reached.} we slightly modify the payoff into
\begin{align}
    q_{\tau}A_{\tau} - \max(F_{min, \tau},X_{\tau})
\end{align}
where $$\tau \in \Tc_{ex} = \left\{ \tau \text{ stopping time } | N_{ex} \le \tau \le  N_{min} + \min \left( C_{\tau}, N_{max} - N_{min} \right) \text{ a.s.} \right\}.$$

Therefore, the trader just solves the following problem\footnote{Once again, we maximize the expected value under $\mathbb{Q}$ because we implicitly assume that the resulting payoff will be $\Delta$-hedged.}
\begin{align}
    \underset{(v_n)_n}{\sup}\ \underset{\tau \in \Tc _{ex}}{\sup} \mathbb E^{\mathbb Q} \left[q_{\tau}A_{\tau} - \max(F_{min, \tau},X_{\tau}) \right].
\end{align}

To solve this problem, we propose in the next section different heuristics for both the execution process $(v_n)_n$ and the stopping time $\tau$.

\section*{Numerical results and discussion}

In this section, we present various numerical examples to illustrate the application of our method. We begin with a very simple buyback contract and then incrementally introduce additional features to demonstrate the method's adaptability and depth.\\

In all the examples, we choose $S_0 = 10$ \euro\  and $\bar \sigma = 0.2\ \text{year}^{-1/2}$.\\

 We use three naive strategies as benchmarks:
 \begin{itemize}
     \item The linear policy: each day, choose $v_n = \frac{F_{max}-X_n}{S_n (T_n-n)}$ with $T_n = \min (N_{min}+C_n, N_{max})$, until the end (no early stopping).
     \item The minmaxtarget policy: each day, choose 
    $$
    v_n = \begin{cases}
        \frac{F_{max}-X_n}{S_n(T_n -n)}\quad \text{if } S_n \le A_n,\\
        \frac{F_{min}-X_n}{S_n (T_n -n)} \quad \text{otherwise,}
    \end{cases}
    $$
    and stop when $S_n<A_n$ and $n\ge N_{ex}$.
    \item The no trade - no early stopping policy: wait until maturity and execute everything at the end.
 \end{itemize}

\subsection*{A simple buyback}

We first consider a buyback with the following parameters:
\begin{itemize}
    \item Maturity: $N:=N_{min}=N_{max}=60\ \text{days}$;
    \item Notional: $F:=F_{min}=F_{max}= 200\text{M}$\euro;
    \item Early exercise: $N_{ex}= 40\ \text{days}$;
    \item Bounds: $v_{min} = 0$, $v_{max} = +\infty$;
    \item Price cap: $S_{max}=+\infty$.
\end{itemize}

We propose a first heuristic strategy (hereafter denoted by Optuna-$\alpha,a$), given as follows:
\begin{itemize}
    \item Each day, choose
$$v_n = \frac{2}{1+e^a} \times \frac{\bar F -X_n}{S_n (T_n-n)},$$
with $\bar F = \frac{F_{min} + F_{max}}{2}.$
\item Stop when $A_n/S_n - 1 \ge \alpha$.
\end{itemize}

We can also try to improve the early stopping side of the strategy, and consider a second heuristic strategy (Optuna-$\alpha,\beta,\gamma,a$) as follows:
\begin{itemize}
    \item Each day, choose
$$v_n = \frac{2}{1+e^a} \times \frac{\bar F -X_n}{S_n (T_n -n)},$$
with $\bar F = \frac{F_{min} + F_{max}}{2}.$
\item Stop when $A_n/S_n - 1 \ge \alpha + \beta \frac{T_n -n}{T_n} + \gamma \frac{X_n}{F_{min}}$.
\end{itemize}

These two strategies are then optimized over $\alpha,\beta,\gamma$, and $a$ using the TPE sampler of the \texttt{optuna} python library and their respective performance are reported in the table below corresponding to a Monte Carlo simulation with 2,000 trajectories.\footnote{Heuristic strategies are optimised upon at the outset and then followed deterministically when we carry out Monte Carlo simulations.} We see in particular that the two strategies clearly outperform the benchmarks in terms of average PnL. We also observe that adding more parameters to the early execution strategy does not improve the performance.
 
\begin{center}
\begin{tabular}{c|c|c|c}

 & PnL in million & Mean (bp) & StdDev (bp) \\
\hline
Linear policy & 0.33 & 16.34 & 15.07 \\
\hline
Minmaxtarget policy & 1.53 & 76.57 & 92.49 \\
\hline
No trade - no early stopping & 0.96 & 47.92 & 573.50 \\
\hline
Optuna-$\alpha$,$a$ & 2.00 & 100.23 & 544.67 \\
\hline
Optuna-$\alpha,\beta,\gamma$,$a$ & 1.98 & 99.19 & 548.31 \\

\end{tabular}
\end{center}

\subsection*{Buyback with cap on the volume}

We now add a cap on the volume that can be traded on each day, and consider a buyback with the following parameters:
\begin{itemize}
    \item Maturity: $N:=N_{min}=N_{max}=60\ \text{days}$;
    \item Notional: $F:=F_{min}=F_{max}= 200\text{M}$\euro;
    \item Early exercise: $N_{ex}= 40\ \text{days}$;
    \item Bounds: $v_{min} = 0$, $v_{max} = 0.2\frac{\bar \Vc}{S_0}$;
    \item Price cap: $S_{max}=+\infty$.
\end{itemize}
Here $\bar \Vc= 400\text{M}$\euro\ denotes the average daily volume.\\

In order to capture the increased complexity of the product, we introduce some new strategies.\\

The third heuristic strategy (Optuna-$\alpha$,$a$,$a_1$) is given by:
\begin{itemize}
    \item Each day, choose
$$v_n = \frac{2}{1+\exp\left( a + a_1 \frac{T_n-n}{T_n} \right)} \times \frac{\bar F -X_n}{S_n (T_n -n)},$$
with $\bar F = \frac{F_{min} + F_{max}}{2}.$
\item Stop when $A_n/S_n - 1 \ge \alpha $.
\end{itemize}

The fourth heuristic strategy (Optuna-$\alpha,a,a_1,a_2$) is given by:
\begin{itemize}
    \item Each day, choose
$$v_n = \frac{2}{1+\exp\left( a + a_1 \frac{T_n-n}{T_n} + a_2 \frac{\max(N_{ex}-n, 0)}{N_{ex}} \right)} \times \frac{\bar F -X_n}{S_n(T_n -n)},$$
with $\bar F = \frac{F_{min} + F_{max}}{2}.$
\item Stop when $A_n/S_n - 1 \ge \alpha $.
\end{itemize}

The fifth heuristic strategy (Optuna-$\alpha,a,a_1,b$) is given by:
\begin{itemize}
    \item Each day, choose
$$v_n = \frac{2}{1+\exp\left( a + a_1 \frac{T_n-n}{T_n} + b \left(\frac AS - 1 \right) \right)} \times \frac{\bar F -X_n}{S_n (T_n -n)},$$
with $\bar F = \frac{F_{min} + F_{max}}{2}.$
\item Stop when $A_n/S_n - 1 \ge \alpha $.
\end{itemize}

And the sixth heuristic strategy (Optuna-$\alpha,a,a_1,b,b_1,c$) is given by:
\begin{itemize}
    \item Each day, choose 
$$v_n = \frac{2}{1+\exp\left( a + a_1 \frac{T_n-n}{T_n} + \left(b + b_1 \frac{T_n-n}{T_n}\right) \left(\frac AS - 1 \right) + c\frac{F_{min}-X_n}{S_n v_{max} (T_n-n)} \right)} \times \frac{\bar F -X_n}{S_n(T_n -n)},$$
with $\bar F = \frac{F_{min} + F_{max}}{2}.$
\item Stop when $A_n/S_n - 1 \ge \alpha $.
\end{itemize}

The results are reported in the table below. Observe in particular that, as soon as we add a cap on the volume, the third benchmark strategy can never finish the program and therefore yields a negative payoff. Again, our optimized heuristic strategies outperform the benchmarks, and increasing the complexity may results in a slightly higher average PnL, but the difference is never significant: simple heuristic strategies seem to be already very close their more complex counterparts.

\begin{center}
\begin{tabular}{c|c|c|c}
& PnL in million & Mean (bp) & StdDev (bp) \\
\hline
Linear policy & 0.33 & 16.34 & 15.07 \\ \hline
Minmaxtarget policy & 1.52 & 76.22 & 91.33 \\ \hline
No trade - no early stopping & -120.00 & -5999.97 & 227.56 \\ \hline
Optuna-$\alpha$,$a$ & 1.62 & 80.87 & 60.34 \\ \hline
Optuna-$\alpha,\beta,\gamma$,a & 1.63 & 81.41 & 63.21\\
\hline
Optuna-$\alpha$,$a,a_1$ & 1.66 & 82.89 & 95.60 \\ \hline 
Optuna-$\alpha,a,a_1,b$ & 1.68 & 84.03 & 115.28 \\ \hline
Optuna-$\alpha,a,a_1,b,b_1,c$ & 1.68 & 83.93 & 119.19 \\
\end{tabular}
\end{center}

\subsection*{Buyback with cap on the volume and floating notional} 

We now consider a feature well-known to practitioners, sometimes called ‘‘flex size’’ or ‘‘Greenshoe’’. More precisely, we consider a buyback program with the following parameters:
\begin{itemize}
    \item Maturity: $N_{min}=N_{max}=60\ \text{days}$;
    \item Notional: $F_{min}= 200\text{M}$\euro, $F_{max} = 250\text{M}$\euro;
    \item Early exercise: $N_{ex}= 40\ \text{days}$;
    \item Bounds: $v_{min} = 0$, $v_{max} = 0.2\frac{\bar \Vc}{S_0\Delta t}$;
    \item Price cap: $S_{max}=+\infty$.
\end{itemize}

We compare the different strategies, and report the results in the table below. Notice first that allowing the trader to buy more shares always results in an increased PnL. Again, our six strategies clearly beat the benchmarks. We also see in this case with floating notional that increasing the complexity of the strategy by adding a dependence of the execution strategy on the $\frac AS$ ratio significantly enhances the average PnL.

\begin{center}
\begin{tabular}{c|c|c|c}
& PnL in million & Mean (bp) & StdDev (bp) \\
\hline
Linear policy & 0.49 & 24.51 & 21.51 \\ \hline
Minmaxtarget policy & 2.21 & 110.66 & 97.32 \\ \hline
No trade - no early stopping & -120.00 & -5999.97 & 227.56 \\ \hline
Optuna-$\alpha_,a$ & 2.42 & 121.06 & 79.34 \\ \hline
Optuna-$\alpha,\beta,\gamma,a$ & 2.45 & 122.52 & 81.04 \\ \hline
Optuna-$\alpha,a,a_1$ & 2.44 & 122.12 & 81.03 \\ \hline
Optuna-$\alpha,a,a_1,a_2$ & 2.45 & 122.31 & 80.54 \\ \hline
Optuna-$\alpha,a,a_1,b$ & 2.75 & 137.68 & 91.71 \\ \hline
Optuna-$\alpha,a,a_1,b,b_1,c$ & 2.80 & 140.13 & 77.02 \\
\end{tabular}
\end{center}

\subsection*{Buyback with cap on the volume, floating notional and price cap} 

Finally, we add a price cap to the program and consider the following parameters:
\begin{itemize}
    \item Maturity: $N_{min}=60\ \text{days}$, $N_{max}=65\ \text{days}$;
    \item Notional: $F_{min}= 200\text{M}$\euro, $F_{max} = 250\text{M}$\euro;
    \item Early exercise: $N_{ex}= 40\ \text{days}$;
    \item Bounds: $v_{min} = 0$, $v_{max} = 0.2\frac{\bar \Vc}{S_0\Delta t}$;
    \item Price cap: $S_{max}=1.2 S_0 = 12$\euro.
\end{itemize}

We compare the different strategies and report the results in the table below.

\begin{center}
\begin{tabular}{c|c|c|c}
& PnL in million & Mean (bp) & StdDev (bp) \\
\hline
Linear policy & 0.24 & 11.75 & 74.57 \\ \hline
Minmaxtarget policy & 2.03 & 101.57 & 118.23 \\ \hline
No trade - no early stopping & -118.77 & -5938.64 & 537.30 \\ \hline
Optuna-$\alpha,a$ & 2.18 & 108.76 & 89.44 \\ \hline
Optuna-$\alpha,\beta,\gamma,a$ & 2.19 & 109.68 & 88.05 \\ \hline
Optuna-$\alpha,a,a_1$ & 2.20 & 109.83 & 85.85 \\ \hline
Optuna-$\alpha,a,a_1,a_2$ & 2.20 & 109.86 & 89.09 \\ \hline
Optuna-$\alpha,a,a_1,b$ & 2.75 & 137.42 & 98.32 \\ \hline
Optuna-$\alpha,a,a_1,b,b_1,c$ & 2.79 & 139.59 & 78.22 \\
\end{tabular}
\end{center}

Again, our six strategies clearly beat the benchmarks. We also see, as above, that increasing the complexity of the strategy by adding a dependence of the execution strategy on the $\frac AS$ ratio significantly enhances the average PnL.

\section*{Concluding remarks}

In conclusion, our research marks a significant shift from traditional approaches that relied on stochastic optimal control tools for pricing and managing share buyback contracts, which were originally inspired by optimal execution methods. By refining and optimizing heuristic strategies, we successfully address the limitations posed by high-dimensional state spaces and the complex selection of risk parameters. Our method simplifies the issue to pricing and hedging a path-dependent payoff, making it tractable with conventional financial tools. This approach not only builds a bridge between theory and application in financial markets but also separates the strategies for delivering shares to the company from the hedging of the payoff, potentially elucidating execution patterns observed in the market.

\section*{Statement and acknowledgment}

This research has been conducted with the support of the Research Initiative ``Modélisation des marchés actions, obligations et dérivés'' financed by HSBC France under the aegis of the Europlace Institute of Finance. Manuel Abellan-Lopez (HSBC), Jérôme Lemue (HSBC) and Stéphane Quentin (HSBC) deserve special thanks for the discussions we had with them. The views expressed are those of the authors and do not necessarily reflect the views or the practices at HSBC.

\bibliographystyle{plain}

\end{document}